\documentclass{aa}
\usepackage{txfonts}
\usepackage{graphicx}
\usepackage{aalongtable}
\usepackage{natbib}
\usepackage{here}
\bibpunct{(}{)}{;}{a}{}{,} % to follow A&A style
\newcommand{\micron}{\mbox{$\mu$m}}
\newcommand{\msun}{M$_{\rm \odot}$}

\newcommand{\ea}{et~al.}
\newcommand{\clumps}{{\tt clumps}}

\begin{document}

\title{The Giant Molecular Cloud associated with RCW~106}
\subtitle{A 1.2~mm continuum mapping study}

\author{B. Mookerjea\inst{1}
  \and C. Kramer\inst{1}
  \and M. Nielbock\inst{2}
  \and L.-$\AA$ Nyman\inst{3}}

\institute{KOSMA, I. Physikalisches Institut, Universit\"at zu
K\"oln, Z\"ulpicher Strasse 77, 50937 K\"oln, Germany
\and Astronomisches Institut der Ruhr-Universit\"at Bochum,
Universit\"atsstrasse 150, 44780 Bochum, Germany
\and Swedish-ESO Submillimetre Telescope, European Southern
Observatory, Casilla 19001, Santiago 19, Chile; Onsala Space
Observatory, 439 92 Onsala, Sweden}

\offprints{B. Mookerjea, \email bhaswati@ph1.uni-koeln.de}

\date{Received /Accepted }

\titlerunning{1.2~mm study of RCW106}
\authorrunning{Mookerjea et al.}

\abstract
{We have mapped the dust continuum emission from the molecular
cloud covering a region of 28~pc$\times$94~pc associated with
the well-known {\sc H~ii} region RCW~106 at 1.2~mm using SIMBA
on SEST. The observations, having an HPBW of 24\arcsec\
(0.4~pc), reveal 95 {\tt clumps}, of which about 50\% have MSX
associations and only 20\% have IRAS associations. Owing
to their higher sensitivity to colder dust and higher angular
resolution the present observations identify new emission
features and also show that most of the IRAS sources in this
region consist of multiple dust emission peaks.  The detected
millimeter sources (MMS) include on one end the exotic MMS5
(associated with IRAS 16183-4958, one of the brightest
infrared sources in our Galaxy) and the bright (and
presumably cold) source MMS54, with no IRAS or MSX
associations on the other end.  Around 10\% of the sources are
associated with signposts of high mass star formation
activity.  Assuming a uniform dust temperature of 20~K we
estimate the total mass of the GMC associated with RCW~106 to
be $\sim 10^5$\,\msun. The constituent millimeter {\tt
clumps} cover a range of masses and radii between 40 to
$10^{4}$~\msun\ and 0.3 to 1.9~pc.  Densities of the clumps
range between (0.5-6) 10$^4$~cm$^{-3}$.  We have decomposed
the continuum emission into {\tt Gaussian} and {\tt arbitrary
shaped  clumps} using the two independent structure analysis
tools {\tt gaussclumps} and {\tt clumpfind} respectively.  The
{\tt clump} mass spectrum was found to have  an index $\alpha$
of $1.6\pm0.3$, independent of the decomposition algorithm
used.  The index of the mass spectrum for the mass and length
scales covered here are  consistent with results derived from
large scale CO observations.  }

\maketitle
\keywords{ISM: clouds \-- ISM: dust, extinction \-- ISM:  {\sc H~ii} regions \--ISM:
structure \-- stars: formation}

\section{Introduction}

Massive stars are believed to form in clusters within molecular
cloud complexes. Owing to their occurrence in more distant crowded
stellar clusters, shorter formation timescales and formation in
regions of high visual extinction, the formation of massive stars
is still poorly understood. Dense, cool molecular gas and dust
cocoons enshroud massive protostars and account for most of the
extinction toward these regions. The extinction is generally so
high that a massive protostar cannot be directly observed even at
the K band (2.2~\micron).  \citet{watson97} succeeded in one of the
rare direct observations of the ionizing star in K band for the
UC{\sc H ii} region G29.96-0.02. Optically thin millimeter and
submillimeter continuum emission from dust cocoons shrouding the
sites of massive star formation is a valuable probe of molecular
cloud structure. Owing to the recent developments in the detection
techniques at millimeter and sub-millimeter wavelengths, large
continuum maps essential for deriving a census of dust emission
peaks arising from the deeply embedded phases of star formation are
becoming available \citep{johnstone2000,motte2001,motte03}.

RCW~106 was discovered by \citet{rodgers60} in a survey of
H$\alpha$ line emission of the southern Galactic plane.  
The Giant Molecular Cloud (GMC) associated with RCW~106, 
was first detected
by \citet{gillespie77} during observations of the molecular
clouds associated with southern Galactic {\sc H ii} regions in
the J=1-0 transition of CO.  Radio continuum mapping
observations at 408~MHz and 5~GHz by \citet{goss70} detected a
number of bright {\sc H ii} regions occurring along a line
almost parallel to the Galactic plane.  IRAS observations have
subsequently shown all of these {\sc H ii} regions to be
bright infrared  sources.  The GMC is located at a distance of
3.6~kpc \citep{lockman79} covering an area of
70\arcmin~$\times$15\arcmin\
%%CKr: 2.6.04: (16\,pc$\times$73\,pc), %CKr.
between $l$=332.5 to 333.7 and appears as a large complex of
bright mid-infrared sources in the Midcourse Space Experiment
(MSX) images between 8 and 21~\micron\ (20\arcsec\
resolution).  The region contains one of the brightest
infrared sources in our Galaxy \citep{becklin73},
IRAS~16183-4958, which harbors the {\sc H~ii} region
G333.6-0.22. In addition to the IRAS maps at the mid and
far-infrared wavebands, far-infrared (FIR) balloon-borne
observations (1\arcmin\ angular resolution) of the dust
continuum at 150 and 210~\micron\ \citep{karnik01} identified
23 emission peaks with dust temperatures between 20 and 40~K.
Largescale CO(1-0) observations by \citet{bronfman89} with an
angular resolution of 9\arcmin\ obtained a global view of the
molecular gas distribution in this region. However most of the
remaining spectroscopic observations toward this GMC are
primarily pointed mode observations of the emissions of the
lines of CO \citep{gillespie77,brand84}, CS
\citep{gardner78,bronfman96}, H$_2$CO \citep{gardner84},
NH$_3$\citep{batchelor77}, H$_2$O, OH and methanol masers
\citep{braz83,caswell97,walsh97} toward predetected {\sc H~ii}
regions and IRAS sources.  These observations suggest ongoing
high-mass star formation activity and enhanced density cores
associated with most of the strong emission peaks.  This
region will also be observed as a part of the Galactic Legacy
Infrared Mid-Plane Survey Extraordinaire (GLIMPSE), a Spitzer
Legacy Science Program \citep{benjamin03} at 3.6, 4.5, 5.8,
and 8.0~\micron\ using the Infrared Array Camera (IRAC).  

In this paper we present the first sensitive, large-scale continuum
survey of the RCW~106 region at millimeter wavelengths. We have
used the 1.2~mm observations to identify $95$ {\tt  clumps}, about
half of which have MSX counterparts. We discuss association of
these {\tt  clumps} with sources detected in MSX and IRAS surveys
in particular.  We have used methods based on {\tt clumpfind}
\citep{williams93,williams94} and {\tt gaussclumps}
\citep{stutzki90} to derive the {\tt clump} mass distribution in
this region.

\section{Observations and data reduction}
The 1.2~mm continuum observations were carried out with the 37
channel bolometer array SIMBA (SEST Imaging Bolometer Array) at
the SEST (Swedish-ESO Submillimeter Telescope) on La Silla, Chile
during the night from July 4 to 5, 2002. A mosaic covering
approximately $1600$\arcsec$\times$5400\arcsec\ ($\alpha\times
\delta$) was observed using the fast scanning method. The mapped region
extends over 28~pc$\times$94~pc. Altogether 8
individual maps were used to construct the large mosaic and the
total observing time was about 9.5 hours.  The zenith opacity
during the observation, measured using skydips, ranged between
0.209 and 0.385. For all but the two submaps at the top, two
coverages were possible, resulting in differences in the noise
levels between the northernmost part of the map and the rest of
it.  The residual noise in the final co-added mosaic is 23
mJy/beam for most of the southern part of the mosaic and is around 50
mJy/beam for the northernmost part.  Uranus was mapped for
calibration purposes and a calibration accuracy of 15\% was
achieved. The HPBW for these observations was 24\arcsec.

\begin{figure*}
\sidecaption
\includegraphics[angle=0,width=14.0cm,angle=0]{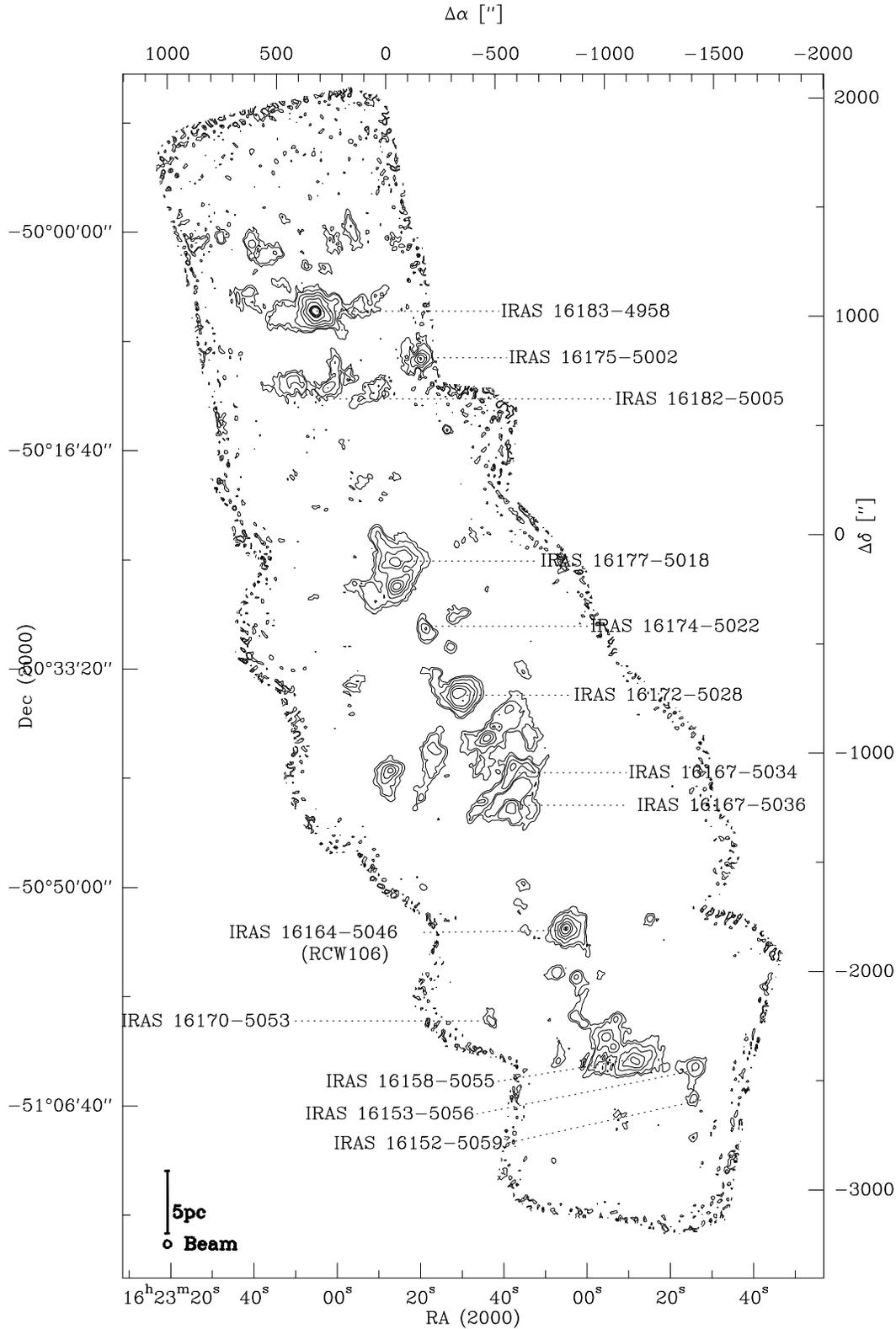}
\caption
{The 1.2~mm mosaic of the star forming regions around RCW~106
observed with SIMBA. The typical 1$\sigma$ rms noise is
23~mJy/beam. The contours correspond to 0.1, 0.2, 0.5,  1, 2, 4, 10, 12, 15,
20, and 40~Jy/beam.  All associated
IRAS point sources are also indicated in the plot.The mapped region
extends over 28~pc$\times$94~pc. Also shown in the plot is the SIMBA beam.
\label{rcw106map}}
\end{figure*}

All data were reduced and analyzed using MOPSI\footnote{MOPSI is a
software package for infrared, millimeter and radio data reduction
developed and constantly upgraded by R. Zylka} according to
the instructions of the SIMBA Observer's Handbook (2002) and
also using methods described by
\citet{chini2003}. Figure~\ref{rcw106map} shows the 1.2~mm dust
continuum emission from the region associated with RCW~106 as
observed by SIMBA, together with the associated IRAS point sources.

\section{Results}

To locate and measure the properties of dust emission peaks or
``{\tt clumps}'' as we shall refer to them in this paper (see
Section 3.4), we have used an automated procedure based on a
two-dimensional variant of the clump-finding algorithm, {\tt
clumpfind} \citep{williams94}.  The {\tt clumps} so identified may
have any arbitrary shape, so that the procedure gives results
similar to what an inspection by eye would have done but being
automated, has far better accuracy. We further imposed the
conditions that in order to be identified as a genuine sub-mm
source, a {\tt clump} must have (1) a deconvolved size larger than
50\% of the HPBW (24\arcsec) and (2) an S/N ratio greater than 5.
Table~\ref{sourcetab} lists the positions, peak intensities,
integrated intensities, derived physical properties (size, mass and
density) of the 95 {\tt clumps} that have been detected at 1.2~mm.

\subsection{Overview of mm sources}

\begin{figure}
\begin{center}
\includegraphics[angle=0,width=8.0cm,angle=0]{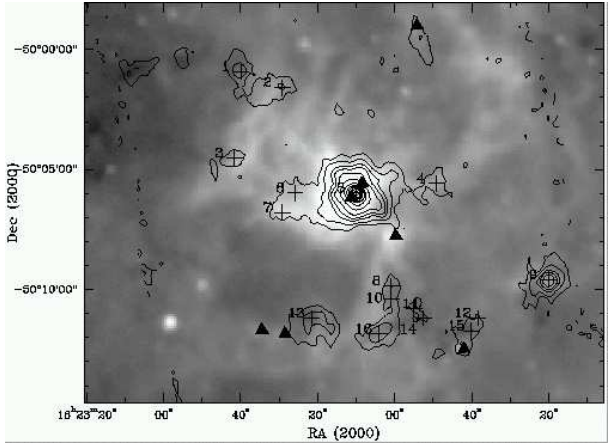}
\caption{Contours of 1.2~mm dust continuum overlayed on the MSX
8~\micron\ image for the  northernmost sub-region. 
%The
%contours drawn are the same multiples of the rms noise of 50
%mJy/beam, as in Figure~\ref{rcw106map}. 
The rms noise for the map is 50~mJy/beam and the contours are 0.2, 0.5, 1.0,
1.5, 2.5, 5.0 Jy/beam  and  10 to 40~Jy/beam in steps of 5. 
The millimeter {\tt clumps} ($+$) and MSX-UC{\sc H~ii} candidates
 (filled triangles) are marked.
\label{msx_simba_comp_1}}
\end{center}
\end{figure}

\begin{figure}
\begin{center}
\includegraphics[angle=0,width=8.0cm,angle=0]{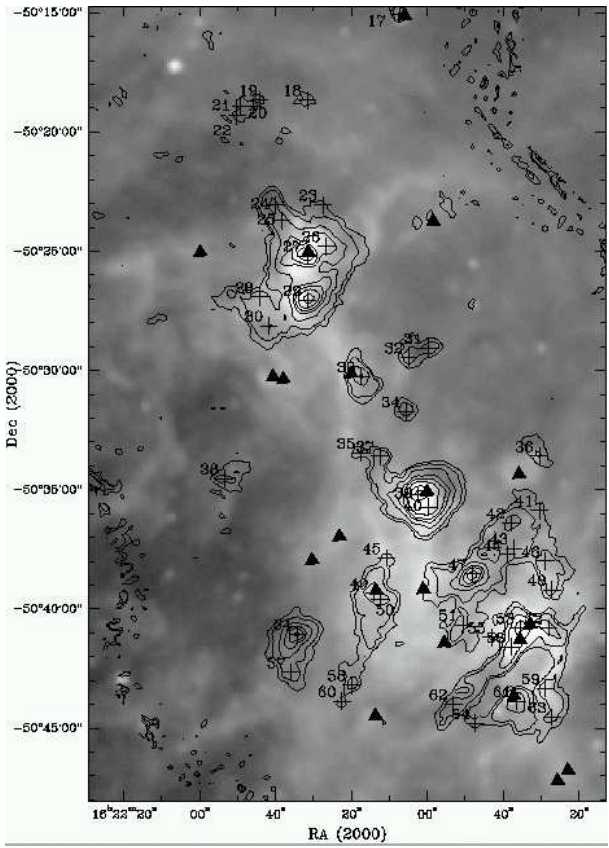}
\caption{Same as in Figure~\ref{msx_simba_comp_1} only for the
middle part of the mapped region.
%The
%contour levels and rms noise are the same as in
%Figure~\ref{rcw106map}.
The contours are at 0.1, 0.2, 0.5, 1, 1.5, 2.5, 5 and 8.5~Jy/beam.
\label{msx_simba_comp_2}}
\end{center}
\end{figure}

\begin{figure}
\begin{center}
\includegraphics[angle=0,width=8.0cm,angle=0]{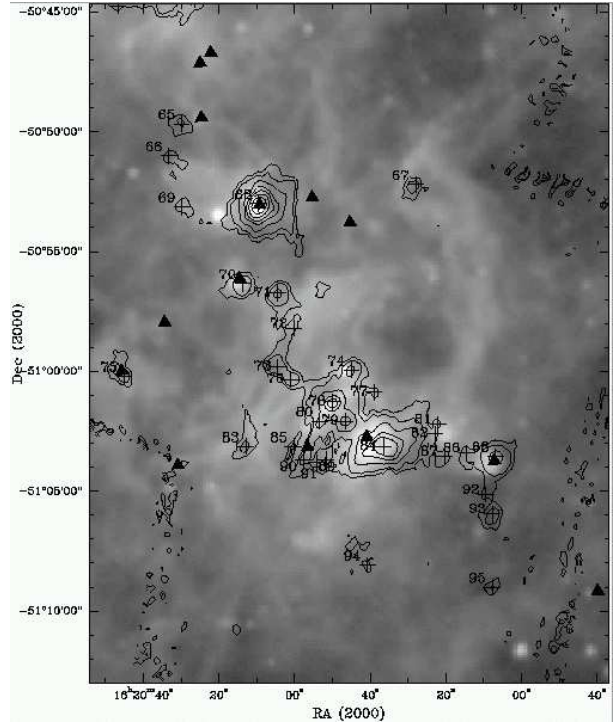}
\caption{Same as in Figure~\ref{msx_simba_comp_2} only for the
southernmost part of the mapped region.
The contours are at for 0.1, 0.2, 0.5, 1, 1.5, 2.5, 5 10 and 12~Jy/beam.
\label{msx_simba_comp_3}}
\end{center}
\end{figure}

To obtain a first estimate of the nature of the sources detected at
1.2~mm we have derived associations of these sources with existing MSX
and IRAS point sources. We have associated a millimeter source with an
MSX (or IRAS or FIR) source if the MSX  (or IRAS or FIR) source lies within a
radius of 40\arcsec\ (90\arcsec\ or 60~\arcsec) from the millimeter
source. We have also identified different signposts of massive star
formation viz.,  {\sc H~ii} regions, masers (H$_2$O, OH, methanol etc)
associated with the {\tt clumps}. In addition, the MSX point sources
for which the logarithm of the ratio of the Band A (8~\micron) to the
Band~E (21~\micron) flux densities, $log(S_8/S_{21})\geq0.57$, are
identified as candidates for UC{\sc H~ii} regions (R. Simon,
$priv.~communication$). This is a direct extension of the
Wood-Churchwell criterion \citep{wood89} for identifying UC{\sc H~ii}
regions based on the the IRAS 12 and 25~\micron\ flux densities.
Figures~\ref{msx_simba_comp_1}, \ref{msx_simba_comp_2} and
\ref{msx_simba_comp_3} show overlays of the MSX (greyscale) 8~\micron\
image with the SIMBA contours, with the mm-sources marked together
with all MSX-UC{\sc H~ii} candidates in the region.  We note that this
method of deriving associations could be severely affected by the
differences in the resolutions and sensitivities of the different
observations under consideration.  Following this method, we identify
two main types of \clumps :

{\bf Type~A :} Clumps with an infrared (IRAS or MSX or FIR)
association.  Table~\ref{typea} lists the 46 type A \clumps\ together
with the associated IRAS and MSX point sources, {\sc H ii} regions,
MSX-UCHII candidates and H$_2$O, methanol or OH masers. Owing to
the higher angular resolution of the 1.2~mm dataset presented here in
contrast to the IRAS beamsizes, on several occasions a single IRAS
source is associated with two or more \clumps.  All sources except
five (MMS40, MMS52, MMS61, MMS63, MMS80) have MSX associations.
Almost all \clumps\ with MSX and IRAS associations are also identified
as candidates for MSX UC{\sc H~ii} regions. Only 7 type~A \clumps\
are found to have  {\sc H~ii} regions associated with them and only 5
of them (MMS5, MMS9, MMS39/40, MMS68 and MMS84) have associated
maser activities.  MMS~54, having only a FIR association, has a methanol
maser associated with it. Type~A sources with associated  {\sc H~ii} regions
are thus the more evolved, massive star-forming \clumps\ of this
region.

{\bf Type~B :} The remaining 49 \clumps\ with no infrared
associations. These are pure sub-mm sources, some of which are
mass or size-wise no different from the Type~A sources (see
Section 3.3).

\subsection{Mass}

Table~\ref{sourcetab} also presents the masses of the \clumps\
($M_{\rm clump}$) derived from their total integrated flux
densities ($F_{\nu}$) using the following relationship:

\begin{equation}
\label{eq-dust-mass}
M_{\rm gas}~=~\frac{F_\nu D^2}{\kappa_\nu B_\nu(T_{\rm dust})}.
\end{equation}

This equation is valid since dust emission at 1.2~mm is optically thin.

In equation\,\ref{eq-dust-mass}, $B_\nu(T_{\rm dust})$ is the Planck
function at the dust temperature $T_{\rm dust}$, $\kappa_{\nu}$ is the
dust mass opacity coefficient, $D$ is the distance.  In the absence of
complementary continuum data, we generally assume a dust temperature
of 20\,K which is typically found in star forming cloud complexes
\citep[e.g.]{johnstone2000,motte03}.  However for the \clumps\ with
associated  {\sc H~ii} regions we have assumed  $T_{\rm dust}$ to be
40~K \citep{karnik01}.  For the dust mass opacity coefficient we adopt
the value of $\kappa_{230GHz}~=~0.005$~cm$^2$~g$^{-1}$
\citet{preibisch93} assuming a gas-to-dust mass ratio of 100, and
$\kappa_\nu$ = $\kappa_{230GHz}$($\nu$/230GHz)$^\beta$, with $\beta$ =
2.  Assuming a uniform dust temperature of 20~K, the total mass of
the region is estimated to be $10^5$\,\msun.

A few additional details about the calculation of the mass:

(1) Given the large population of UC-{\sc H~ii} candidates in
the region, 1.2~mm continuum emission observed with SIMBA is
most likely to be contaminated by the free-free emission
from the {\sc H~ii} regions.  However as mentioned earlier, no
radio continuum image of comparable angular resolution is
available for the entire region.  For the \clumps\ with
associated  {\sc H~ii} regions the contribution of the
free-free emission from the gas nevertheless has been 
estimated using the available 5~GHz flux densities 
\citep{haynes79} and a power law dependence of
S$_\nu\sim\nu^{-0.1}$, for optically thin free-free
emission. The contribution varies between 4 and 36\% (see
Table~\ref{sourcetab}). For the \clumps\ for which the radio
continuum flux densities were not available, despite their
association with an  {\sc H~ii} region, we have assumed a
constant 20\% contribution to the flux densities at 1.2~mm.

(2) Two other sources of uncertainty in converting the
observed flux density at 1.2~mm into gas mass are (a) the dust
to gas mass ratio which typically ranges between 100 and 150
and (b) the dust mass opacity coefficient, which ranges
between 0.003 (diffuse ISM) up to 0.02~cm$^2$~g$^{-1}$
\citep{kruegel94} for dense protostellar environments,
depending on the size and structure of the dust grains. 

%(3) The dust temperatures derived by \citet{karnik01} vary
%between 20 and 40~K.  To take this into account, for the
%\clumps\ with associated  {\sc H~ii} regions, we assume the
%dust temperature to be 40~K, while for the rest of the
%\clumps\ we assume it to be 20~K.

\subsection{Sizes \& densities}

At a distance of 3.6~kpc the HPBW of 24\arcsec\ of SIMBA
corresponds to a diameter of 0.42~pc. The radii of the {\tt clumps}
identified by {\tt clumpfind} range between the resolution limit
to as large as 1.9~pc, with a substantial fraction having radii
larger than 0.6~pc. Thus most of these {\tt clumps} are of the size
of star forming regions, rather than the innermost star
forming cores typically of 0.01 to 0.1~pc size which have
been dealt with in similar mm/sub-mm continuum studies of regions
like $\rho$ Ophiuchi, Orion~B, W~43, IC~5146 etc.
%CKr:
%\citep{johnstone2000,motte1998,motte03,kramer03}.
\citep{motte1998,johnstone2000,motte03,kramer03}.
%CKr.

\begin{figure}
\begin{center}
\includegraphics[angle=0,width=8.0cm,angle=0]{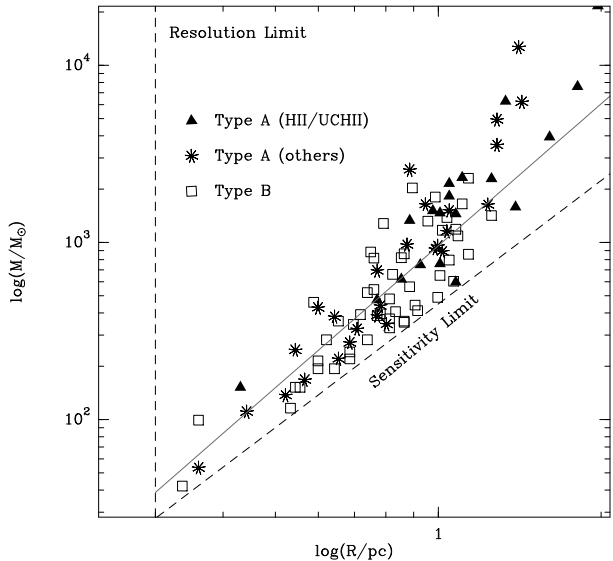}
\caption
{Plot of mass as a function of the 
% CKr: 2.6.04:
% size 
radius (convolved with the beam) 
% CKr.
for all detected {\tt clumps}. The dashed vertical line corresponds to
the minimum size that a clump may have (resolution limit) and the
other dashed line shows the minimum mass that a clump must have in
order to be 1$\sigma$ above the mass detection limit.  The sensitivity
limit corresponds to M$\propto$R$^2$.  Also shown is the
M$\propto$R$^{2.3}$ as found in CO observations
\citep{heithausen1998}, extrapolated to masses higher than 100~\msun.
\label{overview}}
\end{center}
\end{figure}

Figure~\ref{overview} shows  a log-log plot of the estimated
mass as a function of the radii for all the 95 millimeter {\tt
clumps} in this region.  Since the clumps are not point
sources, the completeness limit, measured in total flux or
mass depends on both the range of object sizes and the surface
brightness profiles.  This holds for all surveys that attempt
to measure the integrated flux from extended objects.
Figure~\ref{overview} shows that most of the more massive and
larger \clumps\ also belong to Type~A and are either {\sc
H~ii} regions or candidates for UC-{\sc H~ii} regions.
Figure~\ref{overview} does not show any clear distinction
between the two types of \clumps. There appears to be a
distinct mass-to-radius correlation for all the clumps, although
part of it probably arises from the resolution and sensitivity
limits. The continuous line shows the mass to radius
correlation (M$\propto$R$^{2.3}$) seen in CO clumps
\citep{heithausen1998} for $0.01<R/pc<10$ and
$10^{-4}<$M/\msun$<10^{2}$. For comparison, the Type~A
\clumps\ show a M$\propto$R$^{2.8\pm0.3}$ correlation, whereas
the Type~B \clumps\ exhibit a M$\propto$R$^{2.3\pm0.2}$
correlation. This implies that within the uncertainties the
mass-radius relation for all types of \clumps\ agree well with
the result from \citet{heithausen1998} as well as the
sensitivity limit.

%CKR: 2.6.04
The slope found for the dense, small prestellar objects studied
  by e.g. \citet{motte2001}, is much shallower: $M\propto\,R^{1.1}$ though
  this relation may likewise be affected by the size-dependent
  detection thresholds. However, as pointed out by \citet{motte2001} a
  linear relation is expected for self-gravitating isothermal
  condensations.
%CKr.

Based on the calculated mass and derived sizes, in
Table~\ref{sourcetab} we also present the estimated densities of the
\clumps. We find that the \clumps\ have moderate densities ($\sim
10^4$~cm$^{-3}$) spread over a rather narrow range. The densities are
similar to those observed in the dust continuum study by
\citet{kerton2001} as well as the typical densities found in 
CO emission studies. The
estimated densities also show that the sources under consideration
here are quite different from the high density ($\sim 10^6$~cm$^{-3}$)
condensations which are considered to be prestellar candidates
\citep{motte1998}.

\subsection{Nature of the mm sources}

\citet{blitz93} proposed an operational categorization for structures
of different spatial scales seen in the ISM. In this scheme three
types of structures are defined, viz., clouds, clumps and cores.
Clouds are structures with masses $>10^4$~\msun\ and cores are regions
out of which single stars (or binary systems) form with masses $\leq
10$~\msun\ and sizes $\leq 0.1$~pc. Clumps correspond
to structures with masses and sizes intermediate between clouds and
cores, from which stellar clusters are formed.  The millimeter sources
detected around RCW~106 have radii between 0.3 and 1.9~pc and have
masses between 40 and $10^{4}$~\msun.  Thus we refer to these sources
as \clumps\ and the entire region mapped with a mass of $\sim
10^5$~\msun\ as Giant Molecular Cloud (GMC).  

Given the limitations involved in associating the \clumps\ with the
radio observations detecting various signposts of star forming
activities (viz.,  {\sc H~ii} regions, masers), the evolutionary stages
of the \clumps\ identified here are not accurate.  The case of MMS54
is an apt example.  While its fainter companion MMS57 has an MSX
counterpart, the stronger millimeter source MMS54 has no IRAS or MSX
associations, but was detected both at 150 and 210~\micron\ by
\citet{karnik01} and has an associated methanol maser
\citep[][observed with an angular resolution of
4\arcmin]{ellingsen96}.  This definitely suggests that MMS54 shows
early signs of massive star formation and demands further high angular
resolution sub-mm dust continuum observations to characterize its
nature.  

Broadly, we propose the following evolutionary stages for the \clumps:

(1) Type~A clumps with associated  {\sc H~ii} regions (and some of
them with maser activities as well) are the most evolved high mass
star forming regions.

(2) Type~A clumps which are candidates for MSX-UC{\sc H~ii} regions 
(refer to Section 3.1 for the criterion) are
high mass star forming regions, less evolved than those in (1).

(3) Type~A clumps which have only infrared associations and do not
qualify as UC{\sc H~ii} candidates are possible sites
of low mass star formation at early stages of evolution.

(4) The purely mm clumps (Type B) are cold condensations, which given
their densities are either prestellar or transient
features of  the molecular cloud. 

Most of the existing spectroscopic observations are aimed at the
sources (belonging to category (1) of Type~A) that are bonafide
massive star forming regions. Thus for the newly detected \clumps\
there are no available spectroscopic observations which may be used
to estimate their gravitational stability and hence help to judge
the potential of these \clumps\ to be prestellar objects.

\section{Clump mass distribution}
%{\bf (completely revised)}} 
%% edited by CKr 17.5.04

Molecular cloud structure can be mapped via mm spectroscopy of molecular
lines, continuum (sub)millimeter emission from dust or stellar NIR
absorption by dust. The first method gives kinematical as well as spatial
information and results in a three dimensional cube of data, whereas the
latter two result in two dimensional datasets. We use the second method to
derive information about the structure of the molecular cloud around the
RCW~106 region.
  
For this purpose we have made use of the two most direct structure analysis
tools, viz., {\tt clumpfind} \citep{williams94} and {\tt gaussclumps}
\citep{stutzki90,kramer98} both of which decompose the observed emission
into discrete clumps. The masses derived using {\tt clumpfind} are already
presented in Table~\ref{sourcetab}.  We have also used {\tt
gaussclumps} which follows a completely different clump identification
and extraction algorithm.  In contrast to {\tt clumpfind}, the
{\tt gaussclumps} algorithm iteratively deconvolves the observed emission
into Gaussian shaped {\tt clumps}.  Though this algorithm was originally
written to decompose 3-dimensional molecular line data sets, it can also be
applied to slightly adapted continuum data without modification of the
code.\footnote{We added two 2-D planes to the data set, mimicking two
additional but empty velocity planes bracketing the continuum data set. We
checked that the formal values for the center velocities and the velocity
widths of the {\tt clumps} which the algorithm finds, stay constant, as
expected. Thus, the algorithm effectively works only in the 2-D center plane
containing the continuum data.} \citet{motte03} already used this method to
analyze a continuum data set of W43.
  
For both {\tt clumpfind} and {\tt gaussclumps}, we have selected only those
{\tt clumps} that have deconvolved sizes larger than 50\% of the spatial
resolution of the SIMBA data for further analysis.  This discards the very
small {\tt clumps} which contribute only a tiny fraction to the total mass.

The masses of the identified clumps range between 40 and $10^4$\,\msun\
(Table~\ref{table-clump-masses}).  It is worth noting that particularly for
the low mass clumps there is no one-to-one correspondence between positions
and masses of individual clumps found by {\tt gaussclumps} and {\tt
clumpfind}.  The {\tt gaussclumps} algorithm decomposes a larger fraction of
the total flux into clumps (Table\,\ref{table-clump-masses}) and thus finds
a larger number of clumps.  However we point out here that the purpose of
this structure analysis and construction of the mass spectrum is to
obtain an overview of the statistical properties of the clump
ensemble. Thus, we have used all clumps regardless of their nature
and/or evolutionary stages to build up the mass spectrum.

\begin{figure}
\begin{center}
\includegraphics[angle=0,width=8.0cm,angle=0]{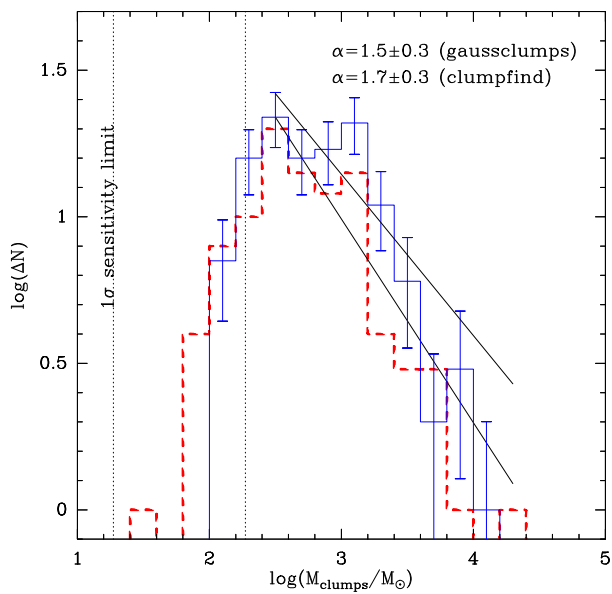}
\caption {\label{fig-mass-spectra}
  The mass spectrum for the clumps identified using {\tt clumpfind}
  (dashed histogram) and {\tt gaussclumps} (drawn histogram)
  respectively. Error bars represent the standard deviation of a
  Poisson distribution $\sqrt{\Delta N}$ and are only drawn for 
  one spectrum for clearness. The dashed vertical line indicates the
  $1\sigma$ (=19\msun) and $10\sigma$ masses given by the average
  noise level of the map of 50\,mJy/beam, the beam size ($24''$,
  i.e. 0.42\,pc), and assuming $T_{\rm dust}=20$\,K.
\label{massspec}}
\end{center}
\end{figure}

Figure\,\ref{fig-mass-spectra} shows the resulting distributions in a
log-log plot of the number of clumps found per mass interval.  We find
that the clump mass distributions have similar shapes.  The slope of
the distribution flattens when going from the clumps of
$\sim10^4$\,\msun\ to clumps of $\sim10^3$\,\msun.  At smaller masses of
$\sim190$\,\msun\ corresponding to the $10\sigma$ mass at
$T_{\rm dust}=20$\,K, the distributions are almost flat. There is a
sharp turnover of the spectra at still lower masses. No clumps are
found beyond the minimum $1\sigma$ mass.  The turnover at masses of
less than the $10\sigma$ mass has often been found in mass spectra
created from CO surveys.  \citet{heithausen1998} showed that this
turnover is not an intrinsic property of the clump distribution but
rather due to the finite resolutions and the finite sensitivity.
Observations at higher resolutions, show a smooth continuation of the
mass spectra without any change in slope.  A linear least squares fit
to the {\tt clump} distribution $dN/dM\propto M^{-\alpha}$ for {\tt
clumps} with $M>10\sigma$, results in linear correlation coefficients
between 80 and 90\% and spectral indices $\alpha$ of between 1.7 and
1.6. Both spectra yield the same index within the standard deviation
of $0.3$.

As pointed out in Section 3.4, star forming {\em cores}
are not detected by our survey. This is an important aspect to keep in
mind while comparing the results of the structure analysis presented
here with the results of 
dust continuum and molecular line studies. 

%CKR.

  Table~\ref{table-clump-masses} compares the properties of the
  clump ensemble obtained for RCW\,106 with those of other dust
  continuum mapping observations studying the fragmentation of the clouds.
  We restrict the table only to those studies which consider
  clumps that are
  more massive than stars and more extended than the
  prestellar cores considered e.g. by \citet{motte1998}.  Interestingly,
  the spectral indices of the clump mass distributions from
  these studies lie at $\sim 1.6$.
  
  This is in contrast to the recent dust continuum maps obtained at
  high resolution for the nearby clouds $\rho$Ophiuchus, Serpens, and
  NGC\,2068/71 \citep{motte1998, testi1998, johnstone2000, motte2001}.
  The properties of the resulting clump distributions have been
  compiled by \citet{motte2001b}.  In these studies, clumps have
  masses down to 0.02\,M$_\odot$, sizes less than 0.05\,pc
  ($\le10\,000$ AU), and typical densities of $10^6$\,cm$^{-3}$.
  These starless and gravitationally bound clumps are therefore good
  candidates for being the progenitors of protostars.  The above
  studies show a common spectral index of the clump mass distribution
  of $\sim2.3$. There are only a few exceptions: the studies of
  \citet{sandell2001} and \citet{coppin2000} show indices of
  $\sim1.5$.
  
  A spectral index of $\sim2.3$ matches with the
  Salpeter IMF \citep{salpeter1955} which is thought to hold
  for stars $>1$M~$_\odot$ \citep{kroupa1993,scalo1998}. There
  are indications in $\rho$Oph that at still lower masses, the
  dust continuum mass spectrum flattens out, as does the
  stellar IMF.  
%CKr.

Interestingly, the mass spectrum found here for
$40<$M/\msun$<10^4$ is similar within the uncertainties to the
mass spectra derived from Galactic large-scale CO maps
covering the mass range $10^{-4}<$M/\msun $<10^4$
%CKr: 2.6.04:
%\citep[e.g.][]{schneider2004,heyer2001,simon2000,kramer98,may1997,
%yonekura1997, dobashi1996,brand1995,solomon1989,casoli1984}
\citep[e.g.][and references
therein]{heyer2001,simon2000,kramer98}
% CKR.
all of which agree with $1.5<\alpha<1.8$. In a recent study of
CO in the Antennae galaxies (NGC 4038/4039),
\citet{wilson2003} found again similar index of 1.4 for a
mass range from $5\,10^6$ to $9\,10^8$\msun\ of the entire cloud
complex using {\tt clumpfind}.

Thus the spectral index of the clump mass distribution found here is
independent of the algorithm used and is consistent with  both dust
continuum and CO mapping studies 
%CKr: 2.6.04:
which probe structures of densities $\sim 10^3-10^4$\,cm$^{-3}$
  as we find in RCW~106. Spectral line
  maps at high spatial resolutions are needed to examine the stability
  of these clumps and find out whether they are gravitationally bound
  or are transient features of the turbulent clouds.
%
% However, the gradual steepening of the slope above the $10\sigma$ mass
% seen in our dust continuum clump mass spectrum for RCW\,106 deviates
% from the linear mass spectra usually derived from CO maps.  It may be
% an intrinsic property of the dust clump distribution.
%CKr.

\section{Summary}

The 1.2~mm dust continuum emission from the GMC associated
with the well-known {\sc H~ii} region RCW~106 has been mapped
at an angular resolution of 24\arcsec\ using SIMBA. In all, 95
mm {\tt clumps} have been identified, more than half of which
are associated with pre-detected IR(MSX or IRAS) point
sources.  The region contains a number of dense massive star
forming clumps which are recognized as candidates for UC{\sc
H~ii} regions and are bright in the far-infrared and
millimeter wavelengths. There exist a large number of
molecular condensations which are only bright in the
millimeter wavelengths and have no IR associations.  MMS54 has
no MSX or IRAS associations but was detected in the FIR by
\citet{karnik01} and has an associated methanol maser; it is a
strong candidate for a star forming region at a very early
stage of evolution.  The continuum emission has been resolved
in to  Gaussian and arbitrary-shaped {\tt clumps} using
the two independent structure decomposition tools {\tt
gaussclumps} and {\tt clumpfind} respectively. The masses of
these {\tt clumps} cover a wide range between 40 to
$10^{4}$~\msun, suggesting that while the entire region is
similar to a GMC, the sources detected are more like star
forming regions (protoclusters) and not cores which would give
rise to individual stars.  The distribution of the {\tt clump}
mass was found to have  a spectral index $\alpha$ of
$1.6\pm0.3$, independent of the decomposition algorithm used.
The spectral index of their distributions match well with
those observed in CO observations covering the same mass
range.

The higher angular resolution of this survey in contrast to
the previous IRAS and far-infrared observations allows the
identification of a number of {\tt clumps} which are
potentially interesting for further detailed continuum
observations in the far-infrared and sub-mm as well as
spectroscopic observations.  Owing to the southern location of
the region, ground-based observations will have to wait for
the upcoming telescopes like APEX and later ALMA.

\acknowledgements

We thank the anonymous referee for many detailed comments which helped to
improve the paper considerably. This project was supported by the {\em
Deutsche Forschungsgemeinschaft} through the grant SFB~494. This research has
made use of the SIMBAD database, operated at CDS, Strasbourg, France and
NASA's Astrophysics Data System Bibliographic Services.

%%%%%%%%%%%%%%%%%%%% Begin References %%%%%%%%%%%%%%%%%%%%%%%%%

\begin{longtable}{lllrrrrc}
\caption{Observed and derived parameters of the millimeter
sources detected in the RCW~106 complex. Dust masses have been
calculated for T$_{\rm dust}$ = 20~K and an emissivity $\beta = 2$.
Column (5) gives the clump radii (convolved with the beam).
{\small 
Estimated free-free contribution to the 1.2~mm flux densities from the
 {\sc H~ii} regions are $^1$ 23\%, $^2$ 20\%, $^3$ 36\%, $^4$ 12\%, $^5$ 4\%, $^6$
18\%. The masses for these sources are calculated assuming a temperature of
40~K.}
\label{sourcetab}} \\
%\small
%\begin{tabular}
\hline\hline
Source&  RA(2000)& Dec(2000)& F$_{peak}$&  Radius &
F$_\nu$ & Mass  & n$_{\mathrm H_2}$\\
& &  & mJy/beam  & pc & Jy & \msun\ & 10$^4$ cm$^{-3}$\\
\hline
\endfirsthead
\caption{continued}\\
\hline\hline
Source&  RA(2000)& Dec(2000)& F$_{peak}$&  Radius &
F$_\nu$ & Mass  & n$_{\mathrm H_2}$\\
& &  & mJy/beam  & pc & Jy & \msun\ & 10$^4$ cm$^{-3}$\\
\hline
\endhead
\hline
\endfoot
MMS1    & 16:22:39.8 & -50:01:00 &   828  &  0.81   &  2.5   &    956  &  1.74 \\ 
MMS2    & 16:22:28.9 & -50:01:41 &   585  &  0.90   &  2.9   &   1097  &  1.45 \\ 
MMS3    & 16:22:41.6 & -50:04:36 &   375  &  0.60   &  1.0   &    376  &  1.68 \\ 
MMS4    & 16:21:49.3 & -50:05:43 &   379  &  0.63   &  1.3   &    500  &  1.93 \\ 
MMS5    & 16:22:10.1 & -50:06:06 & 40013$^1$  &  1.88   & 129.5  &  15936  &  5.44 \\ 
MMS6    & 16:22:25.9 & -50:06:04 &   370  &  0.89   &  2.7   &   1003  &  1.37 \\ 
MMS7    & 16:22:29.2 & -50:06:52 &   430  &  0.89   &  2.2   &    834  &  1.14 \\ 
MMS8    & 16:22:01.1 & -50:09:59 &   502  &  0.51   &  0.7   &    274  &  2.00 \\ 
MMS9    & 16:21:20.3 & -50:09:45 &  4392$^2$  &  0.83   &  8.6   &   1096  &  5.44 \\ 
MMS10   & 16:22:01.2 & -50:10:31 &   475  &  0.47   &  0.8   &    308  &  2.87 \\ 
MMS11   & 16:21:55.4 & -50:10:54 &   356  &  0.49   &  0.5   &    203  &  1.67 \\ 
MMS12   & 16:21:38.8 & -50:11:20 &   258  &  0.47   &  0.4   &    154  &  1.43 \\ 
MMS13   & 16:22:22.1 & -50:11:17 &   775  &  0.65   &  2.5   &    928  &  3.26 \\ 
MMS14   & 16:21:52.9 & -50:11:19 &   273  &  0.64   &  0.8   &    285  &  1.05 \\ 
MMS15   & 16:21:40.5 & -50:11:51 &   546  &  0.93   &  2.8   &   1052  &  1.26 \\ 
MMS16   & 16:22:04.6 & -50:11:58 &  1015  &  0.80   &  3.2   &   1187  &  2.24 \\ 
MMS17   & 16:21:08.0 & -50:15:14 &   578  &  0.32   &  0.3   &    109  &  3.21 \\ 
MMS18   & 16:21:32.3 & -50:18:49 &   195  &  0.51   &  0.4   &    139  &  1.01 \\ 
MMS19   & 16:21:44.9 & -50:18:48 &   183  &  0.41   &  0.2   &     83  &  1.16 \\ 
MMS20   & 16:21:46.6 & -50:19:03 &   178  &  0.24   &  0.1   &     30  &  2.10 \\ 
MMS21   & 16:21:49.9 & -50:19:03 &   161  &  0.55   &  0.4   &    158  &  0.92 \\ 
MMS22   & 16:21:50.8 & -50:19:27 &   198  &  0.72   &  0.7   &    259  &  0.67 \\ 
MMS23   & 16:21:28.4 & -50:23:12 &   163  &  0.47   &  0.4   &    139  &  1.29 \\ 
MMS24   & 16:21:40.9 & -50:23:11 &   564  &  0.62   &  1.6   &    590  &  2.39 \\ 
MMS25   & 16:21:39.3 & -50:23:53 &   761  &  0.84   &  3.5   &   1311  &  2.14 \\ 
MMS26   & 16:21:27.6 & -50:24:56 &  1460$^3$  &  1.25   & 11.2   &   1151  &  1.34 \\ 
MMS27   & 16:21:32.6 & -50:25:20 &  3672$^3$  &  0.96   & 16.5   &   1692  &  4.34 \\ 
MMS28   & 16:21:45.3 & -50:27:04 &   375  &  1.09   &  3.1   &   1183  &  0.88 \\ 
MMS29   & 16:21:32.7 & -50:27:12 &  6502  &  1.27   & 24.7   &   9281  &  4.38 \\ 
MMS30   & 16:21:42.9 & -50:28:15 &   524  &  0.96   &  3.2   &   1198  &  1.31 \\ 
MMS31   & 16:21:00.9 & -50:29:15 &   308  &  0.58   &  0.8   &    282  &  1.40 \\ 
MMS32   & 16:21:06.0 & -50:29:39 &   319  &  0.67   &  0.9   &    346  &  1.11 \\ 
MMS33   & 16:21:18.6 & -50:30:25 &  1095  &  0.90   &  3.5   &   1325  &  1.76 \\ 
MMS34   & 16:21:06.9 & -50:31:55 &   282  &  0.56   &  0.7   &    248  &  1.36 \\ 
MMS35   & 16:21:18.8 & -50:33:37 &   190  &  0.43   &  0.3   &    109  &  1.32 \\ 
MMS36   & 16:20:31.8 & -50:33:48 &   225  &  0.57   &  0.6   &    233  &  1.22 \\ 
MMS37   & 16:21:13.8 & -50:33:45 &   451  &  0.72   &  1.7   &    624  &  1.62 \\ 
MMS38   & 16:21:54.9 & -50:34:40 &   283  &  0.99   &  1.6   &    620  &  0.62 \\ 
MMS39   & 16:21:03.7 & -50:35:23 &  8925$^4$  &  1.19   & 32.6   &   4584  &  6.18 \\ 
MMS40   & 16:21:01.2 & -50:35:54 &  4054$^4$  &  1.14   & 18.3   &   2580  &  3.95 \\ 
MMS41   & 16:20:31.8 & -50:36:04 &   151  &  0.67   &  0.7   &    267  &  0.86 \\ 
MMS42   & 16:20:39.4 & -50:36:36 &   804  &  0.99   &  4.5   &   1675  &  1.67 \\ 
MMS43   & 16:20:38.5 & -50:37:41 &   369  &  0.46   &  0.9   &    331  &  3.29 \\ 
MMS44   & 16:20:40.2 & -50:37:56 &   373  &  0.71   &  1.6   &    593  &  1.60 \\ 
MMS45   & 16:21:12.1 & -50:38:02 &   173  &  0.40   &  0.3   &     98  &  1.48 \\ 
MMS46   & 16:20:30.1 & -50:38:12 &   479  &  0.87   &  2.3   &    849  &  1.25 \\ 
MMS47   & 16:20:49.5 & -50:38:43 &  3322  &  1.29   & 12.1   &   4530  &  2.04 \\ 
MMS48   & 16:20:28.4 & -50:39:24 &   298  &  0.74   &  1.1   &    406  &  0.97 \\ 
MMS49   & 16:21:15.6 & -50:39:29 &   763  &  0.74   &  2.6   &    965  &  2.30 \\ 
MMS50   & 16:21:13.9 & -50:39:45 &   762  &  1.11   &  4.4   &   1664  &  1.18 \\ 
MMS51   & 16:20:52.1 & -50:40:51 &   348  &  1.11   &  2.7   &   1029  &  0.73 \\ 
MMS52   & 16:20:29.4 & -50:41:00 &   741$^2$  &  0.93   &  3.4   &    431  &  1.52 \\ 
MMS53   & 16:20:36.9 & -50:41:00 &  1268$^2$  &  0.71   &  3.5   &    448  &  3.55 \\ 
MMS54   & 16:21:35.8 & -50:41:11 &  2765  &  1.14   &  9.6   &   3598  &  2.35 \\ 
MMS55   & 16:20:44.5 & -50:41:23 &   173  &  0.55   &  0.5   &    173  &  1.00 \\ 
MMS56   & 16:20:39.4 & -50:41:48 &   968$^2$  &  0.86   &  4.3   &    548  &  2.45 \\ 
MMS57   & 16:21:37.6 & -50:42:49 &   359  &  0.85   &  1.8   &    691  &  1.09 \\ 
MMS58   & 16:21:21.6 & -50:43:22 &   326  &  0.90   &  1.5   &    575  &  0.76 \\ 
MMS59   & 16:20:30.3 & -50:43:33 &   326  &  0.84   &  1.8   &    665  &  1.08 \\ 
MMS60   & 16:21:24.2 & -50:44:01 &   153  &  0.26   &  0.1   &     38  &  2.09 \\ 
MMS61   & 16:20:37.8 & -50:44:04 &   893  &  0.90   &  4.2   &   1563  &  2.07 \\ 
MMS62   & 16:20:54.7 & -50:44:11 &   426  &  0.93   &  2.3   &    860  &  1.03 \\ 
MMS63   & 16:20:28.6 & -50:44:44 &   258  &  0.87   &  1.7   &    642  &  0.94 \\ 
MMS64   & 16:20:48.8 & -50:44:59 &   160  &  0.60   &  0.5   &    203  &  0.91 \\ 
MMS65   & 16:20:32.0 & -50:49:56 &   233  &  0.69   &  0.8   &    293  &  0.86 \\ 
MMS66   & 16:20:35.4 & -50:51:17 &   159  &  0.67   &  0.6   &    237  &  0.76 \\ 
MMS67   & 16:19:30.4 & -50:52:32 &   328  &  0.63   &  0.7   &    278  &  1.07 \\ 
MMS68   & 16:20:11.9 & -50:53:17 & 12460$^5$  &  1.70   & 36.1   &   5548  &  2.56 \\ 
MMS69   & 16:20:32.1 & -50:53:24 &   170  &  0.72   &  0.7   &    255  &  0.66 \\ 
MMS70   & 16:20:16.1 & -50:56:38 &   482  &  0.78   &  1.4   &    541  &  1.10 \\ 
MMS71   & 16:20:06.8 & -50:57:02 &   603  &  0.94   &  2.1   &    789  &  0.92 \\ 
MMS72   & 16:20:02.6 & -50:58:30 &   155  &  0.92   &  1.2   &    436  &  0.54 \\ 
MMS73   & 16:20:06.8 & -51:00:06 &   439  &  0.62   &  1.0   &    391  &  1.58 \\ 
MMS74   & 16:19:47.4 & -51:00:15 &   662  &  0.73   &  1.9   &    702  &  1.74 \\ 
MMS75   & 16:20:47.6 & -51:00:27 &   417  &  0.63   &  0.9   &    342  &  1.32 \\ 
MMS76   & 16:20:03.5 & -51:00:38 &   424  &  0.52   &  0.7   &    259  &  1.78 \\ 
MMS77   & 16:19:41.4 & -51:01:11 &   176  &  0.77   &  0.8   &    297  &  0.63 \\ 
MMS78   & 16:19:52.5 & -51:01:35 &  1905  &  0.74   &  5.0   &   1867  &  4.45 \\ 
MMS79   & 16:19:49.1 & -51:02:24 &  1331  &  0.75   &  3.9   &   1476  &  3.38 \\ 
MMS80   & 16:19:55.8 & -51:02:23 &   670  &  0.61   &  1.7   &    639  &  2.72 \\ 
MMS81   & 16:19:24.4 & -51:02:33 &   154  &  0.66   &  0.7   &    248  &  0.83 \\ 
MMS82   & 16:19:25.3 & -51:02:56 &   164  &  0.44   &  0.3   &    120  &  1.36 \\ 
MMS83   & 16:20:15.4 & -51:03:25 &   263  &  0.86   &  1.2   &    470  &  0.71 \\ 
MMS84   & 16:19:38.9 & -51:03:28 &  2499$^6$  &  1.48   & 21.9   &   2871  &  2.45 \\ 
MMS85   & 16:20:02.7 & -51:03:26 &   247  &  0.33   &  0.2   &     79  &  2.12 \\ 
MMS86   & 16:19:16.7 & -51:03:45 &   195  &  0.55   &  0.5   &    195  &  1.13 \\ 
MMS87   & 16:19:22.8 & -51:03:53 &   158  &  0.76   &  0.8   &    319  &  0.70 \\ 
MMS88   & 16:19:09.2 & -51:03:53 &   746  &  0.86   &  2.8   &   1063  &  1.61 \\ 
MMS89   & 16:19:54.2 & -51:03:59 &   529  &  0.68   &  1.3   &    477  &  1.47 \\ 
MMS90   & 16:20:00.1 & -51:03:58 &   300  &  0.42   &  0.5   &    177  &  2.31 \\ 
MMS91   & 16:19:56.7 & -51:04:06 &   258  &  0.26   &  0.2   &     71  &  3.90 \\ 
MMS92   & 16:19:11.7 & -51:05:30 &   159  &  0.52   &  0.4   &    158  &  1.09 \\ 
MMS93   & 16:19:09.9 & -51:06:17 &   315  &  0.64   &  0.8   &    316  &  1.16 \\ 
MMS94   & 16:19:43.1 & -51:08:23 &   155  &  0.85   &  0.9   &    353  &  0.56 \\ 
MMS95   & 16:19:10.0 & -51:09:21 &   224  &  0.42   &  0.3   &    109  &  1.42 \\ 
%\hline
%\end{tabular}
\end{longtable}

\begin{table*}[h]
\caption{Type~A : \clumps\ with Infrared associations
\label{typea}}
\begin{tabular}{lccccc}
\hline
Source & IRAS & MSX &  {\sc H~ii} & UC{\sc H~ii} & Masers\\
 & & &  & candidates & \\
\hline
\hline
MMS2    & \ldots&   333.6896-00.2066  & \ldots  & N &\ldots \\
\ldots   & \ldots&   333.6931-00.2023  &\ldots   & N  &\ldots\\
MMS4    & \ldots&   333.5684-00.1648  & \ldots  & N  &\ldots\\
\ldots   & \ldots&   333.5681-00.1608  &\ldots   & N  &\ldots\\
MMS5    &  16183-4958     &    333.6044-00.2165   &
 333.6-0.22  & Y & H$_2$O, OH\\
\ldots   &  \ldots         &    333.6046-00.2124   & \ldots &
Y&\ldots\\
\ldots   &  \ldots         &    333.6056-00.2044   & \ldots &
\ldots&\ldots\\
MMS7    &\ldots &   333.6316-00.2562  & \ldots  & N  &\ldots\\
MMS8    &\ldots &   333.5497-00.2455  &\ldots   & N  &\ldots\\
MMS9    &  16175-5002     &    333.4678-00.1591   &
 333.46-0.16 & N & CH$_3$OH, OH\\
MMS10   &\ldots &   333.5360-00.2565  &\ldots   & N  &\ldots\\
MMS15   &\ldots &   333.4749-00.2362  & \ldots  & Y  &\ldots\\
MMS16   &  16182-5005     &    333.5328-00.2670   & \ldots & N&\ldots\\
MMS17   &\ldots &   333.3755-00.2007  & \ldots  & Y &\ldots \\
MMS26   &  16177-5018     &    333.3080-00.3655   &
 333.3-0.4 & Y&\ldots\\
MMS27   &  16177-5018     &    333.3080-00.3655   & 
 333.3-0.4 & Y&\ldots\\
MMS28   &\ldots &   333.3113-00.4254  &   & N &\ldots \\
MMS29   &\ldots &   333.2898-00.3898  &   & N &\ldots \\
MMS33   &  16174-5022     &    333.2271-00.4053   & \ldots &
Y&\ldots\\
\ldots   &  \ldots         &    333.2276-00.4084   & \ldots  &
N&\ldots\\
MMS36   &\ldots &   333.0929-00.3539  &   & N  &\ldots\\
MMS39   &  16172-5028     &    333.1313-00.4268   &
 333.1-0.4 & Y & CH$_3$OH\\
MMS40   &  16172-5028     &   \ldots              & 
 333.1-0.4 & \ldots & CH$_3$OH\\
MMS45   &\ldots &   333.1241-00.4756  &\ldots &   N &\ldots \\
MMS47   &\ldots &   333.0688-00.4445  &\ldots &   N  &\ldots\\
MMS49   & \ldots&   333.1082-00.5011  &\ldots  & Y &\ldots \\
MMS50   & \ldots&   333.1082-00.5011  & \ldots & Y &\ldots \\
MMS52   &  16167-5034     &   \ldots              &
 333.0-0.4  & \ldots&\ldots\\
MMS53   &  16167-5034     &    333.0149-00.4431   &
 333.0-0.4 & Y&\ldots\\
MMS54   &  S19$^\ast$     &                       &  \ldots & N   & CH$_3$OH  \\
%\ldots   &  \ldots         &   G333.0125-00.4555   & \ldots\\
MMS56   &  16167-5034     &    333.0125-00.4555   &
 333.0-0.4 & Y&\ldots \\
MMS57   &\ldots &   333.1166-00.5816  &\ldots  &  N  &\ldots\\
MMS59   &  16167-5036     &   \ldots              & \ldots &
\ldots&\ldots\\  
MMS60   &\ldots &   333.0675-00.5855  & \ldots &  N  \\  
MMS61   &  16167-5036     &    332.9881-00.4864   & \ldots &
Y&\ldots\\
MMS63   &  16167-5036     &   \ldots              & \ldots  &
\ldots&\ldots\\
MMS67   &  16156-5046     &   \ldots              & \ldots &
\ldots&\ldots\\
MMS68   &  16164-5046     &    332.8269-00.5489   &
 332.8-0.6 & Y & H$_2$O\\
MMS70   &\ldots &   332.8007-00.5953  &\ldots  &  Y &\ldots \\
MMS74   &\ldots &   332.7024-00.5866  & \ldots &  N &\ldots \\
MMS75   &  16170-5053     &    332.8137-00.6980   & \ldots &
Y&\ldots\\
MMS78   &\ldots &   332.6926-00.6121  &\ldots &   N &\ldots \\
MMS81   & \ldots &   332.6213-00.5701  &\ldots &  N&\ldots \\
MMS82   & \ldots &   332.6213-00.5701  &\ldots &  N&\ldots \\
MMS84   &  16158-5055     &    332.6555-00.6129   &
 332.7-0.6 & Y&CH$_3$OH,H$_2$O\\
MMS85   &\ldots &   332.6853-00.6463  &\ldots &  N &\ldots \\
MMS90   &\ldots &   332.6853-00.6463  &\ldots &  N  &\ldots\\
MMS86   &  16153-5056     &   \ldots              & \ldots &
\ldots &\ldots\\
MMS88   &  16153-5056     &    332.5860-00.5634   & \ldots &
Y &\ldots\\  
MMS92   &  16152-5059     &   \ldots              & \ldots  &
\ldots &\ldots\\
MMS93   &  16152-5059     &    332.5600-00.5902   & \ldots &
N &\ldots\\
\hline
\hline
\end{tabular}

{\small $^\ast$ FIR detection at 150 and 210~\micron\ by \citet{karnik01}.}
\end{table*}

\begin{table*}
\caption{Comparison of the masses of the \clumps\ identified in
RCW~106 with large scale dust continuum studies of molecular clouds
finding clumps or fragments more massive than 30\,M$_\odot$ and sizes
of more than 0.05\,pc.  $N_{\rm cl}$ s the number of {\tt clumps}
identified. Sizes refer to deconvolved FWHM sizes. $M_{\rm tot}^{\rm
clumps}$ is the total mass of all {\tt clumps}. $\alpha$ is the
spectral index of the mass distribution shown in
Figure\,\ref{fig-mass-spectra} and  $n$ is the density in cm$^{-3}$.
\label{table-clump-masses}}
%\label{tab-dust-literature}}}
\small
\begin{tabular}{lrrrrrrrrl}
\hline\hline
Cloud & Distance & HPBW      & $N_{\rm cl}$ & Sizes & Mass Range  & $M_{\rm tot}^{\rm clumps}$     & $\alpha$    & log(n) & Reference \\
      & [pc]     & [$''$/pc] &     & [pc]  & [M$_\odot$] &
      \msun\       & & log[cm$^{-3}$] & \\
\hline
RCW\,106  & 3600 & 24/0.4    & 95  & 0.2--3.7 & 37--16000 & 1.0 10$^5$ & $1.7\pm0.3$ & 3.7--4.5 & this paper using {\tt clumpfind} \\ 
RCW~106  & 3600  & 24/0.4    & 122   & 0.2--3.7  & 110-14000 &1.5 10$^5$ &  $1.5\pm0.3$  & \ldots & this paper using {\tt
gaussclumps}\\
KR\,140   & 2300 &  14.5/0.16& 22  &  0.2--0.7  & 0.5--130 & 590 &  $1.5\pm0.04$  & 3.1--5.1 & \citet{kerton2001} \\
M\,8      & 1700 & 12/$\sim0.1$ & 75  & \ldots     & 0.3--43  & 280 &  $1.7\pm0.6$   &\ldots & \citet{tothill2002} \\
\hline
\end{tabular}
\end{table*}
\end{document}